\documentclass[useAMS,usenatbib]{mn2e}
\usepackage{graphicx}
\usepackage{txfonts}
\usepackage{tabularx}
\usepackage{multirow}
\title[Vacuum-UV absorption spectroscopy of interstellar ice analogs. Isotopic effects.]{Vacuum-UV absorption spectroscopy of interstellar ice analogs. Isotopic effects.}
\author[G. A. Cruz-Diaz, G. M. Mu\~noz Caro, and Y.-J. Chen]{G. A. Cruz-Diaz$^{1}$\thanks{E-mail: cruzdga@cab.inta-csic.es}, G. M. Mu\~noz Caro$^{1}$, and Y.-J. Chen$^{2,3}$\\ 
$^{1}$Carretera de Ajalvir, km 4, Torrejon de Ardoz, 28850 Madrid, Spain\\
$^{2}$Space Sciences Center and Department of Physics and Astronomy,\\
\hspace{0.02cm}
University of Southern California, Los Angeles, CA 90089-1341, USA\\
$^{3}$Department of Physics, National Central University, Jhongli City, Taoyuan Country 32054, Taiwan}
\begin{document}

\date{Accepted 2013 --------. Received 2013 --------; in original form 2013 --------}

\pagerange{\pageref{firstpage}--\pageref{lastpage}} \pubyear{2013}

\maketitle

\label{firstpage}

\begin{abstract}
This paper reports the first measurements of solid-phase vacuum-ultraviolet (VUV) absorption cross sections of heavy isotopologues present in icy dust grain mantles of dense 
interstellar clouds and cold circumstellar environments. Pure ices composed of D$_2$O, CD$_{3}$OD, $^{13}$CO$_{2}$, and $^{15}$N$^{15}$N were deposited at 8 K, a value similar 
to the coldest dust temperatures in space. The column density of the ice samples was measured \emph{in situ} by infrared spectroscopy in transmittance. VUV spectra of the ice samples 
were collected in the 120-160 nm (10.33-7.74 eV) range using a commercial microwave discharged hydrogen flow lamp as the VUV source. Prior to this work, we have recently 
submitted a similar study of the light isotopologues (Cruz-Diaz et al. 2013a; Cruz-Diaz et al. 2013b). The VUV 
spectra are compared to those of the light isotopologues in the solid phase, and to the gas phase spectra of the same molecules. Our study is expected to improve very 
significantly the models that estimate the VUV absorption of ice mantles in space, which have often used the available gas phase data as an approximation of the absorption cross 
sections of the molecular ice components. We will show that this work has also important implications for the estimation of the photodesorption rates per absorbed photon in the ice.
\end{abstract}


\section{Introduction}


After molecular hydrogen (H$_2$), the molecules H$_2$O, CO, CO$_2$, and CH$_3$OH, are among the most abundant in the interstellar medium, as it has been inferred from observations of 
the gas and solid phase (Mumma \& Charnley 2011, and references therein). The main elements, and their corresponding isotopes, which compose most volatile molecules in the 
interstellar medium, are H:D, $^{12}$C:$^{13}$C, $^{14}$N:$^{15}$N, and $^{16}$O:$^{17}$O:$^{18}$O. 

Deuterium enrichment can be the result of low temperature gas-grain reactions because of the differences in zero-point energies between deuterated and non-deuterated species (Wilson et al. 1973). 
Observation toward prestellar cores indicates, in gas phase, that abundances of singly deuterated molecules are typically higher than the cosmic atomic D/H ratio of 1.5 $\times$ 10$^{-5}$(Linsky 2003), 
also, doubly and triply deuterated molecules have been observed with D/H ratios reaching $\sim$ 30 \% for D$_2$CO and $\sim$ 3 \% for CD$_3$OH (see, Ceccarelli et al. 1998; Loinard et al. 2002; 
Parise et al. 2004; Ratajczak et al. 2009). Deuterated methanol molecules were detected in the gas phase toward low-mass class 0 protostars with abundances up to about 60 \% relative 
to CH$_3$OH (Parise et al. 2006). D$_2$O has been detected toward the solar-type protostar IRAS 16293-2422 (Butner et al. 2007; Vastel et al. 2010). \cite{Roberts} showed that the multiply 
deuterated isotopologues of H$_3$$^+$ can efficiently transfer deuterons to other neutral molecules in very cold ($\leq$ 20 K) gas depleted of its CO (because the CO molecules are frozen 
onto refractory dust grain mantles). 

Isotopic substitution often alters the chemical and physical properties of atoms and molecules, resulting in differences in absorption spectra and reaction rates. Therefore, 
measurements of the isotopic compositions of various species can be used to interpret the physico-chemical histories and the chemical reaction pathways in these environments. 
In particular, isotope effects in the non-dissociative photoionization region of molecular nitrogen play an important role in isotopic fractionation in planetary 
atmospheres and other environments (e.g., interstellar molecular clouds, the solar nebula, and in the atmospheres of Earth, Mars, and Titan) in which 
N$_2$ and VUV radiation are present (see, Croteau et al. 2011, and references therein). 

Carbon dioxide is an important constituent of quiescent and star forming molecular clouds (Gerakines et al. 1999, and references therein). It is primarily present in the 
solid state (van Dishoeck et al. 1996). The $^{13}$CO$_2$ isotope has been detected with a two orders of magnitude lower abundance with respect to CO$_2$ (d'Hendecourt et al. 1996). 
The stretching band of $^{13}$CO$_2$ is an independent and sensitive probe of the ice mantle composition (Boogert et al. 2000, and references therein). 
Studies in the gas phase have shown that the $^{12}$C/$^{13}$C ratio increases with Galacto-centric radius (Wilson \& Rood 1994; Keene et al. 1998). In the solid phase, the behavior 
of this ratio agrees with the gas phase studies (Boogert et al. 2000). Also, determination of this ratio is an important input for evolutionary models of the Galaxy, since $^{12}$C 
is produced by Helium burning by massive stars, which can be converted to $^{13}$C in the CNO cycle of low- and intermediate-mass stars at later times (Boogert et al. 2000).

The average cross section for a certain spectral range are useful when there is no information (flux, photodesorption rate) for each specific wavelength within that range. 
An example are the photodesorption rate values reported in the literature (e.g., \"Oberg et al. 2007, 2009; Mu\~noz Caro et al. 2010), which correspond to the full continuum 
emission spectrum of the hydrogen VUV lamps (an analog of the secondary UV emission in dense clouds). To estimate the photodesorption rate per absorbed photon in that case, 
what is used is the average photon energy and the average VUV absorption cross section in the same range. In addition, these average VUV absorption cross sections allow 
comparison with previous works that estimated those average values in similar spectral ranges in an indirect way (i.e., with no use of VUV spectroscopy), see e.g, 
Cottin et al. (2003). 

Heavy isotopologues, in the context of laboratory astrophysics, are often used to study ice photoprocessing and, in particular ice photodesorption experimentally, 
to avoid problems with contamination (e.g., $^{13}$CO, $^{15}$N$_2$; see Oberg et al. 2007, 2009; Fayolle et al. 2013). 
In addition, it is interesting to search for differences in the absorption of these ices compared to those made of light isotopologues. 

The estimation of the VUV-absorption cross sections of molecular ice components allows to calculate the photon absorption of icy grains in that range. 
In addition, the VUV-absorption spectrum as a function of photon wavelength is required to study the photo-desorption processes over the full photon emission energy range. 

It is therefore important to study the physical and chemical properties of molecules containing heavy isotopes. This study focusses on the isotopic effects observed in the 
vacuum-ultraviolet (VUV) absorption spectra of three of the most abundant inter- and circumstellar species in the solid phase: H$_2$O, CH$_3$OH, and CO$_2$. Among the possible 
isotopologues, the fully deuterated ones were selected, D$_2$O and CD$_3$OD, in addition to $^{13}$CO$_2$. Also $^{15}$N$^{15}$N, henceforth abbreviated as $^{15}$N$_2$, was 
included in our study to explore the isotopic effects in a homonuclear diatomic molecule that was also observed in space (Bergin et al. 2002; Belloche \& Andr\'e 2003). 
Two recent papers report the VUV absorption cross sections of the light isotopologues in the ice (Cruz-Diaz et al. 2013a, 2013b), henceforth referred to as Papers I and II, 
respectively). The data in Paper I and II were used for comparison to this work. In addition, the VUV absorption spectra of the same molecules in the gas phase were also adapted 
to illustrate the differences between the gas and the solid phase samples.

\section{Experimental protocol}

The measurements were conducted using the Interstellar Astrochemistry Chamber (ISAC). This set-up and the standard experimental protocol were described in detail in \cite{Munoz}. 
ISAC mainly consists of an ultra-high-vacuum (UHV) 
chamber, with pressure typically in the range P = 3-4.0 $\times$ 10$^{-11}$ mbar at room temperature, where an ice layer is made by deposition of a gas species onto a cold finger at 8 K. The low 
temperature is achieved by means of a closed-cycle helium cryostat. The ice sample can be either UV-irradiated or warmed up to room temperature. The evolution of the solid sample 
was monitored with \emph{in situ} Fourier transform infrared (FTIR) spectroscopy in transmittance and VUV spectroscopy. The chemical compounds used for the experiments described 
in this paper were: D$_2$O(liquid), Cambridge Isotope Laboratories, Inc (C.I.L.) 99.9\%; CD$_3$OD(liquid), C.I.L. 99.8\%; $^{13}$CO$_2$(gas), C.I.L. 99.0\%; and $^{15}$N$_2$(gas), 
C.I.L. 98.0\%. 

The deposited ice layer was VUV irradiated using a microwave discharged hydrogen flow lamp (MDHL), from Opthos Instruments. The source has a UV-flux of 
$\approx 2 \times 10^{14}$ cm$^{-2}$ s$^{-1}$ at the sample position, measured by CO$_{2}$ $\to$ CO actinometry, see \cite{Munoz}. The Evenson cavity of 
the MDHL is refrigerated with air. The VUV spectrum was measured routinely {\em in situ} during the experiments with the use of a McPherson 0.2 meter focal length VUV 
monochromator (Model 234/302) with a photomultiplier tube (PMT) detector equipped with a sodium salicylate window, optimized to operate from 100-500 nm (11.27-2.47 eV), with a 
spectral resolution of 0.4 nm. The characterization of the MDHL spectrum was previously reported (Chen et al. 2010; Paper I) and was discussed in more 
detail by \cite{Chen2}. 

\section{VUV spectroscopy}
\label{VUV}

VUV absorption cross sections were obtained for pure ices composed of D$_{2}$O, CD$_{3}$OD, $^{13}$CO$_2$, and $^{15}$N$_{2}$. These measurements were performed following the 
procedure described in Paper I and summarized below. The column density of the deposited ice was calculated using FTIR spectroscopy in transmittance, according to the formula

\begin{equation}
N = \frac{1}{\mathcal{A}} \int_{band} \tau_{\nu}d{\nu}
\label{1}
\end{equation}

where $N$ is the column density of the ice, $\tau_{\nu}$ the optical depth of the band, $d\nu$ the wavenumber differential in cm$^{-1}$, and $\mathcal{A}$ is the band strength in 
cm molecule$^{-1}$. The VUV absorption cross section was estimated according to the Beer-Lambert law

\begin{equation}
I_t(\lambda)=I_0(\lambda) {e}^{-\sigma(\lambda) N}
\label{2}
\end{equation}

where $I_{t}(\lambda)$ is the transmitted intensity for a given wavelength $\lambda$, $I_{0}(\lambda)$ the incident intensity, $N$ is the column density in cm$^{-2}$ obtained using 
eq.~\ref{1}, and $\sigma$ is the VUV absorption cross section in cm$^{2}$. 

For each ice spectrum a series of three measurements was performed: i) the emission spectrum of the VUV-lamp was measured, to 
monitor the intensity of the main emission bands, ii) the emission spectrum transmitted by the MgF$_{2}$ substrate window was measured, 
to monitor its transmittance, and iii) the emission spectrum transmitted by the substrate window with  
the deposited ice on top was measured. The absorption spectrum of the ice corresponds to the spectrum of the substrate with the ice 
after subtraction of the bare MgF$_2$ substrate spectrum.

A priori, the VUV absorption cross section of the ice was not known. Therefore, several measurements for different values of the initial ice column density were performed to improve 
the spectroscopy. Table ~\ref{table1} provides the infrared band positions and band strengths of D$_{2}$O, CD$_{3}$OD, and $^{13}$CO$_2$ used to estimate the column density. 
Solid $^{15}$N$_{2}$ does not display absorption features in the mid-infrared, therefore the column density of $^{15}$N$_{2}$ was thus measured using the expression

\begin{equation}
N = \frac{\rho_{N_2} \; d_H}{N_A \; m_{N_2}}
\label{3}
\end{equation}

where $\rho_{N_2}$ is the density of the $N_2$ ice, see Table~\ref{table1}, $m_{N_2}$ is the molar mass of the $N_2$ molecule, $N_A$ is the Avogadro constant (6.022 $\times$ 10$^{23}$ mol$^{-1}$), 
and $d_H$ is the ice thickness in cm. The latter was estimated following the classical interfringe relation 

\begin{equation}
d_H= \frac{1}{2n_H \Delta \nu}
\label{4}
\end{equation}

where $n_H$ is the refractive index of the ice at deposition temperature, and $\Delta \nu$ is the wavenumber difference between two adjacent 
maxima or minima of the fringes observed in the infrared spectrum of the ice.

\begin{table}
\centering
\caption{Infrared band positions, infrared band strengths ($\mathcal{A}$), column density ($N$) in ML (as in previous works, one ML is here defined as 10$^{15}$ molecules cm$^{-2}$), 
and refractive index ($n_H$) of the samples used in this work. Pure $^{15}$N$_{2}$ ice does not display any features in the mid-infrared.}
\begin{tabular}{ccccc}
\hline
\hline
\footnotesize{Species}&\footnotesize{ Position}&\footnotesize{ $\mathcal{A}$}&\footnotesize{ N}&\footnotesize{ $\rho$}\\
&\footnotesize{ [cm$^{-1}$]}&\footnotesize{ [cm molec$^{-1}$]}&\footnotesize{ [ML]}&\footnotesize{ gr cm$^{-3}$}\\
\hline
\footnotesize{ D$_{2}$O}&\footnotesize{ 2413 }&\footnotesize{ 1.0 $\pm$ 0.2 $\times10^{-16 \quad a}$ }&\footnotesize{ 266 $\pm$ 10 }&\footnotesize{ 1.05}\\
\footnotesize{ CD$_{3}$OD}&\footnotesize{ 973 }&\footnotesize{  7.0 $\pm$ 0.3 $\times10^{-18 \quad b}$ }&\footnotesize{ 65 $\pm$ 8 }&\footnotesize{ 1.14}\\
\footnotesize{ $^{13}$CO$_2$}&\footnotesize{ 2276 }&\footnotesize{ 7.8 $\pm$ 0.1 $\times10^{-17 \quad c}$ }&\footnotesize{ 321 $\pm$ 12}&\footnotesize{ --}\\
\footnotesize{ $^{15}$N$_{2}$}&\footnotesize{ -- }&\footnotesize{ -- }&\footnotesize{ 4009 $\pm$ 410}&\footnotesize{ 0.94}\\
\hline
\end{tabular}
\\
{\small $^{a,b}$ Calculated by the us, see Section \ref{VUV}, $^c$ \cite{Gerakines}}\\
\label{table1}
\end{table}

No IR band strength values were found in the literature for the D$_2$O and CD$_3$OD species. These values were therefore calculated using eqs.~\ref{4},~\ref{3}, and~\ref{1}. 
Refractive indices of solid H$_2$O, CH$_3$OH, and N$_2$ were used as an approximation (1.30, 1.39, and 1.21, respectively, see Mason et al. 2006, Hudgins et al. 1993, and Satorre et al. 2008). 
Error values for the column density in Table ~\ref{table1} 
have been estimated taking into account the error in the calculation of the column density and the column density decrease by UV irradiation during the VUV spectral acquisition.

The main emission peaks of the MDHL fall at 121.6 nm (Lyman-$\alpha$), 157.8 nm, and 160.8 nm (molecular H$_2$ bands). These peaks are thus also present in the secondary VUV photon 
spectrum generated by cosmic rays in dense interstellar clouds and circumstellar regions where molecular hydrogen is abundant (Gredel et al. 1989). For this reason, the 
VUV absorption cross section values measured at these wavelengths are provided for each molecule in the following sections.  

The VUV absorption cross section spectra of D$_2$O, CD$_3$OD, $^{13}$CO$_2$, and 
$^{15}$N$_2$ ices were fitted using the sum of two or more Gaussian profiles using an in-house IDL code. These fits correspond to the lowest $\chi^2$ values. 
Table~\ref{tableGauss} summarizes the Gaussian profile parameters used to fit the spectra of the 
different ice compositions deposited at 8 K.

\begin{table}
\centering
\caption{Gaussian parameter values used to fit the spectra of the different molecular ices deposited at 8 K.}
\begin{tabular}{cccc}
\hline
\hline
{\small Molecule}&{\small Centre}&{\small FWHM}&{\small Area}\\
&{\small [nm]}&{\small [nm]}&{\small [$\times$ 10$^{-17}$ cm$^{2}$ nm]}\\
\hline
&&&\\
D$_{2}$O&$\sim$120.0&17.6&7.9\\
&141.5&16.2&9.5\\
&151.2&9.9&1.2\\
&&&\\
CD$_{3}$OD&$\sim$120.2&25.9&26.7\\
&145.7&20.9&11.3\\
&160.5&11.5&1.4\\
&&&\\
$^{13}$CO$_{2}$&115.3&4.2&1.8\\
&126.4&9.9&2.1\\
&&&\\
$^{15}$N$_{2}$&115.5&0.5&0.8\\
&117.0&1.1&1.5\\
&119.2&1.1&1.7\\
&120.8&1.1&2.8\\
&123.0&1.1&3.4\\
&123.5&1.6&1.0\\
&125.0&0.7&3.2\\
&126.1&1.6&0.6\\
&127.4&0.7&3.4\\
&128.5&0.9&1.3\\
&129.9&0.8&1.9\\
&130.8&0.8&2.4\\
&132.1&0.7&0.7\\
&133.2&0.8&3.2\\
&134.8&1.1&0.3\\
&136.2&0.5&4.9\\
&138.0&2.1&0.1\\
&139.0&0.8&2.4\\
&142.2&0.6&4.6\\
&145.4&1.8&3.8\\
\hline
\end{tabular}
\label{tableGauss}
\end{table}

\subsection{Solid deuterium oxide}

\begin{figure}
\centering
\includegraphics[width=\columnwidth]{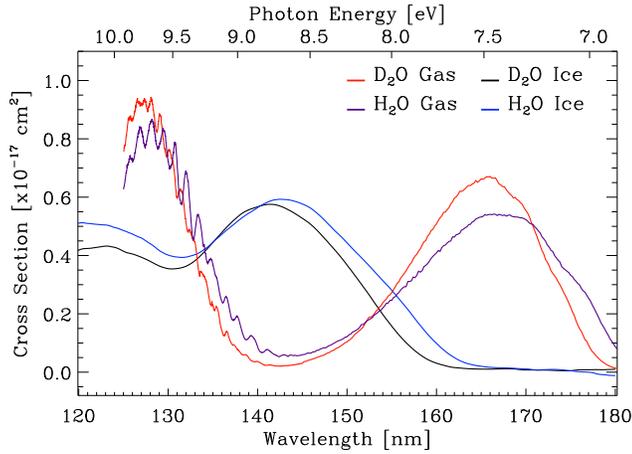}
\caption{VUV absorption cross section as a function of photon wavelength (bottom X-axis) and photon energy (top X-axis) of D$_{2}$O ice deposited at 8 K, black trace. Blue trace is 
the VUV absorption cross section spectrum of solid phase H$_{2}$O, adapted from Paper I. Red and violet traces are the VUV absorption cross section spectra of gas phase D$_{2}$O 
and H$_2$O, respectively, adapted from Cheng et al. (2004) and Chung et al. (2001).}
\label{D2O}
\end{figure}

The VUV absorption cross section spectrum of D$_{2}$O ice (black trace) and H$_2$O ice (blue trace) are displayed in Fig.~\ref{D2O}. \cite{Cheng1} and \cite{Chung} report 
the VUV absorption cross sections of D$_2$O and H$_2$O in the gas phase, depicted in Fig.~\ref{D2O} as red and violet traces, respectively. The 
transition 4a$_{1}$:\~A$^{1}$B$_{1} \leftarrow $ 1b$_{1}$:\~X$^{1}$A$_{1}$ accounts for the absorption in the 145-180 nm region, which reaches its maximum at 166.0 nm for D$_2$O and 
at 167.0 nm for H$_2$O in the gas phase, this accounting for a shift of $\sim$ 1 nm. The same transition was observed for both solid D$_2$O and H$_2$O, with bands centered at 
141.4 nm and 142.6 nm, respectively. This corresponds to a shift of 24.6 $\pm$ 0.4 nm for D$_2$O and 24.4 $\pm$ 0.4 nm for H$_2$O ices compared to the gas phase. Solid D$_2$O 
presents a maximum in the VUV absorption cross section with a value of 5.8$^{+0.2}_{-0.2}$ $\times$ 10$^{-18}$ cm$^{-2}$, a value close to the one estimated for solid H$_2$O, 
6.0$^{+0.4}_{-0.4}$ $\times$ 10$^{-18}$ cm$^{-2}$. The portion of the band in the 120-132 nm range (attributed to the transition \~B$^{1}$A$_{1}$ $\leftarrow$ \~X$^{1}$A$_{1}$, 
according to \cite{Lu}) is present in the four spectra, but due to the MgF$_2$ window cutoff in our set-up it was not possible to determine the position of the maximum for this 
band in the solid samples. 

The average VUV absorption cross section of solid D$_2$O has a value of 2.7$^{+0.1}_{-0.1}$ $\times$ 10$^{-18}$ cm$^{2}$ in the 120-165 nm (10.35-7.51 eV) spectral region, i.e. 
lower than the 3.4$^{+0.2}_{-0.2}$ $\times$ 10$^{-18}$ cm$^{2}$ value of solid H$_2$O. The total integrated VUV absorption cross section of solid D$_2$O is 1.2$^{+0.3}_{-0.3}$ 
$\times$ 10$^{-16}$ cm$^{2}$ nm (8.6$^{+0.1}_{-0.1}$ $\times$ 10$^{-18}$ cm$^{2}$ eV) in the same spectral region, which again is low compared to solid H$_2$O, 1.8$^{+0.1}_{-0.1}$ 
$\times$ 10$^{-16}$ cm$^{2}$ nm. The VUV absorption cross sections of D$_{2}$O ice at 121.6 nm, 157.8 nm, and 160.8 nm are, respectively, 4.4$^{+0.1}_{-0.1}$ $\times$ 10$^{-18}$ 
cm$^{2}$, 0.8$^{+0.1}_{-0.1}$ $\times$ 10$^{-18}$ cm$^{2}$, and 0.3$^{+0.05}_{-0.05}$ $\times$ 10$^{-18}$ cm$^{2}$, i.e. lower than the values for H$_2$O, respectively, 5.2$^{+0.4}_{-0.4}$ 
$\times$ 10$^{-18}$ cm$^{2}$, 1.7$^{+0.1}_{-0.1}$ $\times$ 10$^{-18}$ cm$^{2}$, and 0.7$^{+0.05}_{-0.05}$ $\times$ 10$^{-18}$ cm$^{2}$. The VUV absorption cross section of D$_{2}$O 
in the gas phase has an average value of 3.4 $\times$ 10$^{-18}$ cm$^{2}$. D$_{2}$O gas data have been also integrated in the 120-180 nm range, giving a value of 1.9 $\times$ 
10$^{-16}$ cm$^{2}$ nm (1.1 $\times$ 10$^{-17}$ cm$^{2}$ eV). Both of them, the average and the integrated values, are larger than the ones obtained for solid D$_{2}$O. The 
VUV absorption cross sections of gas phase D$_{2}$O at 157.8 nm and 160.8 nm are, respectively, 4.0 $\times$ 10$^{-18}$ cm$^{2}$ and 5.5 $\times$ 10$^{-18}$ cm$^{2}$, also 
larger than the solid phase measurements. No gas phase data was found for the Ly-$\alpha$ wavelength (121.6 nm).

\subsection{Solid deuterated methanol}

\begin{figure}
\centering
\includegraphics[width=\columnwidth]{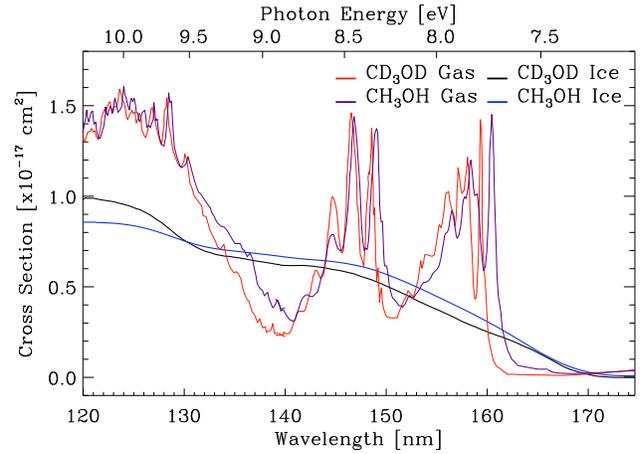}
\caption{VUV absorption cross section as a function of photon wavelength (bottom X-axis) and UV-photon energy (top X-axis) of CD$_{3}$OD ice deposited at 8 K, black trace. Blue 
trace is the VUV absorption cross section spectrum of solid phase CH$_{3}$OH adapted from Paper I. Red and violet traces are the VUV absorption cross section spectra of gas phase 
CD$_{3}$OD and CH$_3$OH, respectively, adapted from Cheng et al. (2002). }
\label{CD3OD}
\end{figure}

Fig.~\ref{CD3OD} shows the VUV absorption cross section of solid CD$_{3}$OD, black trace, and solid CH$_3$OH, blue trace, as a function of wavelength and photon energy. \cite{Cheng2} 
reported the VUV absorption cross section spectra of CD$_3$OD (red trace) and CH$_3$OH (violet trace) in the gas phase, see Fig.~\ref{CD3OD}. The VUV spectra 
corresponding to the gas phase contain plenty of features, while solid VUV spectra are very smooth, with no distinct local maxima. Paper I reports a bump centered at 146.9 nm 
(associated to the 2$^{1}$A'' $\leftarrow$ X$^{1}$A' molecular transition) for solid CH$_3$OH; this band is centered at 145.7 nm for solid CD$_{3}$OD. These maxima were estimated 
using Gaussian fits of the bands. The peaks centered at 146.5 nm and 159.3 nm for gas phase CD$_3$OD are shifted to shorter wavelengths with respect to gas phase CH$_3$OH (peaks centered at 146.8 nm 
and 160.4 nm, respectively). The MgF$_2$ window cutoff in our setup, near 114 nm, only allowed the detecting of a fraction of the broad band corresponding to the 
3$^{1}$A'' $\leftarrow$ X$^{1}$A' molecular transition. This band is present in the four spectra of Fig.~\ref{CD3OD}.

The average VUV absorption cross section of solid CD$_3$OD has a value of 4.6$^{+0.2}_{-0.4}$ $\times$ 10$^{-18}$ cm$^{2}$ in the 120-175 nm (10.33-7.04 eV) spectral region, quite 
similar to the solid CH$_3$OH value of 4.4$^{+0.4}_{-0.7}$ $\times$ 10$^{-18}$ cm$^{2}$. The total integrated VUV absorption cross section of solid CD$_3$OD is 2.6$^{+0.1}_{-0.3}$ 
$\times$ 10$^{-16}$ cm$^{2}$ nm (1.7$^{+0.1}_{-0.2}$ $\times$ 10$^{-17}$ cm$^{2}$ eV) in the same spectral region, very close to the solid CH$_3$OH value, 2.7$^{+0.2}_{-0.4}$ 
$\times$ 10$^{-16}$ cm$^{2}$ nm, reported in Paper I. At the Ly-$\alpha$ wavelength, 121.6 nm, the VUV absorption cross section of CD$_3$OD ice is higher than the value 
corresponding to CH$_3$OH ice (9.7$^{+0.8}_{-1.1}$ $\times$ 10$^{-18}$ cm$^{2}$ and 8.6$^{+0.7}_{-1.3}$ $\times$ 10$^{-18}$ cm$^{2}$, respectively). For the H$_2$ molecular 
transitions at 157.8 nm and 160.8 nm, the VUV absorption cross sections of CD$_3$OD ice (2.9$^{+0.2}_{0.6}$ $\times$ 10$^{-18}$ cm$^{2}$ and 2.2$^{+0.2}_{0.5}$ $\times$ 10$^{-18}$ 
cm$^{2}$) are lower than the values corresponding to CH$_3$OH ice (3.8$^{+0.3}_{-0.6}$ $\times$ 10$^{-18}$ cm$^{2}$ and 2.9$^{+0.2}_{-0.4}$ $\times$ 10$^{-18}$ cm$^{2}$). The VUV 
absorption cross section of CD$_{3}$OD in the gas phase has an average value of 8.6 $\times$ 10$^{-18}$ cm$^{2}$, almost twice larger than the value measured for the solid phase. 
CD$_{3}$OD gas data were integrated in the 120-175 nm range giving a value of 3.4 $\times$ 10$^{-16}$ cm$^{2}$ nm (2.2 $\times$ 10$^{-17}$ cm$^{2}$ eV), i.e, larger than the VUV 
absorption cross section (2.6 $\times$ 10$^{-16}$ cm$^{2}$ nm) of solid CD$_{3}$OD. The VUV absorption cross sections of CD$_{3}$OD gas at 121.6 nm, 157.8, and 160.8 nm are, 
respectively, 13.4 $\times$ 10$^{-18}$ cm$^{2}$, 10.8 $\times$ 10$^{-18}$ cm$^{2}$, and 0.6 $\times$ 10$^{-18}$ cm$^{2}$, which are also larger than the ice phase measurements 
provided above, except for the 160.8 nm value.

\subsection{Solid carbon-13C dioxide}

\begin{figure}
\centering
\includegraphics[width=\columnwidth]{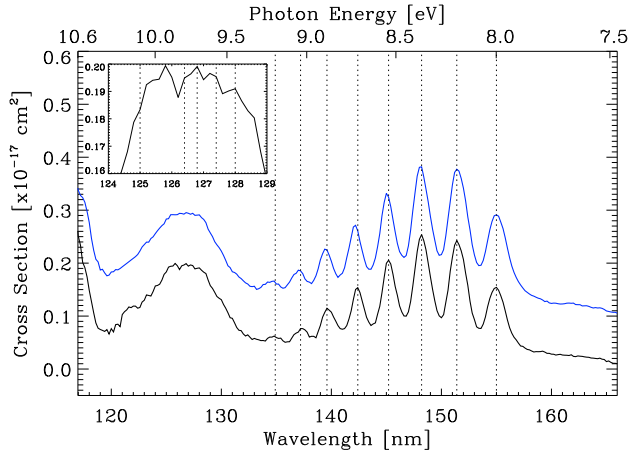}
\caption{VUV absorption cross section as a function of photon wavelength (bottom X-axis) and VUV photon energy (top X-axis) of $^{13}$CO$_{2}$ ice deposited at 8 K, black trace. 
Blue trace is the VUV absorption cross section spectrum of solid CO$_{2}$ adapted from Paper II. Inlet figure is a close-up of the $^{13}$CO$_{2}$ VUV absorption cross section in the 
124-129 nm range. The spectrum of solid CO$_2$ was offset by 1 $\times$ 10$^{-18}$ cm$^{2}$ for clarity.}
\label{13CO2}
\end{figure}

The VUV absorption cross section of $^{13}$CO$_{2}$ ice as a function of the wavelength and photon energy is shown in Fig.~\ref{13CO2}, black trace. It is similar to the one 
reported for CO$_2$ ice in Paper II, depicted as a blue trace in Fig.~\ref{13CO2}. A broad band centered at 9.8 eV, assigned to the $^{1}\Pi_{g} \leftarrow ^{1}\Sigma^{+}_{g}$ transition, is 
observed in these spectra. Paper II reports a vibrational structure in the 120.0-133.0 nm range for CO$_2$; these weak features were poorly resolved in our spectrometer. 
Fig.~\ref{13CO2} inlet shows the same bands for $^{13}$CO$_2$ ice. For comparison, the dotted lines in the inlet represent the five band positions reported in Paper II for solid 
CO$_2$. 

All the discrete bands observed beyond 130 nm correspond to the absorption of photo-produced CO in the CO$_2$ ice matrix. It was therefore not possible to measure the 
spectrum of pure CO$_2$ ice with our experimental configuration. For this, a synchrotron radiation source is required, see Paper II and ref. therein. Similarly, the bands of 
photo-produced $^{13}$CO are present in the spectrum of $^{13}$CO$_2$ ice, see Fig.~\ref{13CO2}. The proportion of $^{13}$CO relative to the deposited $^{13}$CO$_2$ is around 
14\% in this experiment, as it was inferred from integration of the infrared absorption features of $^{13}$CO$_2$ at 2283 cm$^{-1}$ and $^{13}$CO at 2092 cm$^{-1}$. Features 
centered at 155.0 nm, 151.4 nm, 148.2 nm, 145.2 nm, 142.4 nm, 139.6 nm, 137.2 nm, and 135.0 nm correspond to the photoproduced $^{13}$CO. The measured VUV spectrum corresponds 
therefore to a mixture of $^{13}$CO$_{2}$ and $^{13}$CO. An important effect is the shift of the $^{13}$CO features in this experiment with respect to those of pure CO ice, but 
the spectrum of $^{13}$CO was not available for comparison to our results. This issue was discussed in Paper II for CO in the CO$_2$ ice matrix, which was compared to pure CO ice. Such 
ice mixture effects have important implications for the VUV absorption of ice in space, where the molecular components are either mixed or layered in the ice mantles. 

Upper limits for the average and the total integrated VUV absorption cross sections were calculated after subtraction of the $^{13}$CO spectrum; they are, respectively, 
6.9 $^{+0.6}_{-1.0}$ $\times$ 10$^{-19}$ cm$^{2}$ and 3.1$^{+0.3}_{-0.4}$  $\times$ 10$^{-17}$ cm$^{2}$ nm (2.3$^{+0.1}_{-0.2}$ $\times$ 10$^{-18}$ cm$^{2}$ eV). These values are 
comparable with the CO$_2$ values reported in Paper II, 6.7$^{+0.5}_{-0.9}$ $\times$ 10$^{-19}$ cm$^{2}$ and 2.6$^{+0.2}_{-0.3}$ $\times$ 10$^{-17}$ cm$^{2}$ nm for the average and 
the total integrated VUV absorption cross sections, respectively. The VUV absorption cross section of $^{13}$CO$_{2}$ ice at 121.6 nm is 1.1$^{+0.2}_{-0.3}$ $\times$ 10$^{-18}$ 
cm$^{2}$, very close to the 1.0$^{+0.1}_{-0.2}$ $\times$ 10$^{-18}$ cm$^{2}$ value for CO$_2$. No previous gas or solid phase VUV spectra of $^{13}$CO$_2$ were found in the 
literature.

\subsection{Solid nitrogen-$^{15}$N$_2$}

\begin{figure}
\centering
\includegraphics[width=\columnwidth]{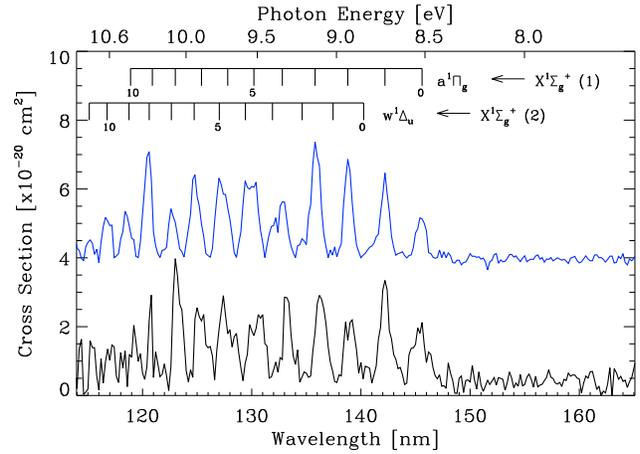}
\caption{VUV absorption cross section as a function of photon wavelength (bottom X-axis) and VUV photon energy (top X-axis) of $^{15}$N$_{2}$ ice deposited at 8 K, black trace. 
Blue trace is the VUV absorption cross section spectrum of solid phase N$_{2}$ adapted from Paper II. The spectrum of solid N$_2$ was offset by 4 $\times$ 10$^{-20}$ cm$^{2}$ 
for clarity.}
\label{15N2}
\end{figure}

The VUV absorption cross section of $^{15}$N$_2$ ice, black trace in Fig.~\ref{15N2}, analogous to that of N$_2$, blue trace, is very low. For this reason, a deposition of about 
$N$ = 4009 $\pm$ 410 $\times$ 10$^{15}$ molecules cm$^{-2}$ was required to detect the absorption features. Paper II summarizes the complete study of solid and gas phase N$_2$. Solid 
$^{15}$N$_2$ should present the same vibrational structure as solid N$_2$ in the 114-147 nm (10.87-8.43 eV) region. The two systems (attributed to a$^{1}\Pi_{g}$ $\leftarrow$ 
X$^{1}\Sigma_{g}^{+}$ and w$^{1}\Delta_{u}$ $\leftarrow$ X$^{1}\Sigma_{g}^{+}$ transitions) can be appreciated in Fig.~\ref{15N2}. The noise level in these measurements was high 
compared to the other ices studied, due to the low intensity of the bands and the detection limit of the VUV spectrometer. Some features are shifted to a shorter wavelengths above the 0.4 nm resolution 
of our measurements, but these shifts did not exceed 0.8 nm.

The average VUV absorption cross section of $^{15}$N$_2$ ice has a value of 8.7 $\times$ 10$^{-21}$ cm$^{2}$, i.e. higher than 7.0 $\times$ 10$^{-21}$ cm$^{2}$ 
calculated for N$_2$ in Paper II. The total integrated VUV absorption cross section of $^{15}$N$_2$ ice has a value of 3.0 $\times$ 10$^{-19}$ cm$^{2}$ nm (2.2 $\times$ 10$^{-20}$ 
cm$^{2}$ eV) in the 114.6-146.8 nm (10.82-8.44 eV) spectral region, which can be compared to 2.3 $\times$ 10$^{-19}$ cm$^{2}$ nm for N$_2$ ice. 
The VUV-absorption cross section spectrum of $^{15}$N$_{2}$ is not as well resolved as in other works but this should not affect the VUV-absorption cross section scale that we 
provide, because no integration of the band area is involved, see Paper II. The VUV absorption cross section at Ly-$\alpha$ (121.6 nm) is very low, we estimated an upper limit value 
of 1.5 $\times$ 10$^{-21}$ cm$^{2}$. There are no observable VUV absorption features at the molecular hydrogen band wavelengths (157.8 and 160.8 nm). No gas or solid 
phase VUV spectra of $^{15}$N$_2$ were found in the literature. 

\section{Astrophysical implications and final conclusions}

The absorption of energetic photons by gas phase molecules and dust grains in various space environments is a key issue in astrophysics. If the absorption cross sections are known 
for the photon wavelength range of interest, a quantitative estimation of the photon absorption and the photon penetration depth in the absorbing material can be attained. 

The absorbing ice column density of a species in the solid phase, can be calculated from the VUV 
absorption cross section following

\begin{equation}
N(\lambda)= \frac{- 1}{\sigma(\lambda)} \ln \left( \frac{I_t(\lambda)}{I_0(\lambda)} \right) 
\end{equation}

where $I_{t}(\lambda)$ is the transmitted intensity for a given wavelength $\lambda$, $I_{0}(\lambda)$ the incident intensity, $N(\lambda)$ is 
the absorbing column density in cm$^{-2}$, and $\sigma(\lambda)$ is the cross section in cm$^{2}$. 
Table ~\ref{penetration} summarizes the absorbing column densities of the ice species for an absorbed photon flux of 95\% and 99\% using the cross 
section value at Ly-$\alpha$, the average cross section in the 120-160 nm range, and the maximum cross section in the same range. The values corresponding to the lighter 
isotopologues are reported in Papers I and II.

\begin{table}
\centering
\caption{Absorbing column densities, of the different ice species, corresponding to an absorbed 
photon flux of 95\% and 99\%. ``Ly-$\alpha$'' corresponds to the cross section at the Ly-$\alpha$ wavelength, 121.6 nm.
``Avg.'' corresponds to the average 
cross section in the 120-160 nm range. ``Max.'' corresponds to the maximum cross section in the same wavelength range.}
\begin{tabular}{ccccccc}
\hline
\hline
&\multicolumn{3}{c}{{\small 95\% photon absorption}}&\multicolumn{3}{c}{{\small 99\% photon absorption}}\\
{\small species}&{\small Ly-$\alpha$}&{\small Avg.}&{\small Max.}&{\small Ly-$\alpha$}&{\small Avg.}&{\small Max.}\\
&\multicolumn{3}{c}{{\small ($\times$10$^{17}$ molecule cm$^{-2}$)}}&\multicolumn{3}{c}{{\small ($\times$10$^{17}$ molecule cm$^{-2}$)}}\\
\hline
{\small D$_2$O}&{\small 6.8}&{\small 11.1}&{\small 5.3}&{\small 10.5}&{\small 17.1}&{\small 8.1}\\
{\small CD$_{3}$OD}&{\small 3.1}&{\small 6.5}&{\small 3.1}{\small }&{\small 4.7}&{\small 10.0}&{\small 4.7}\\
{\small $^{13}$CO$_2$}&{\small 27.2}&{\small 43.7}&{\small 12.0}&{\small 41.8}&{\small 67.1}&{\small 18.4}\\
{\small $^{15}$N$_2$}&{\small 19971}&{\small 3443}&{\small 749}&{\small 30701}&{\small 5293}&{\small 1151}\\
\hline
{\small H$_{2}$O}&{\small 5.8}&{\small 8.3}&{\small 4.9}&{\small 8.9}&{\small 13.0}&{\small 7.7}\\
{\small CH$_3$OH}&{\small 3.5}&{\small 5.7}&{\small 3.4}&{\small 5.4}&{\small 8.7}&{\small 5.3}\\
{\small CO$_2$}&{\small 29.3}&{\small 44.5}&{\small 15.1}&{\small 45.1}&{\small 68.4}&{\small 23.3}\\
{\small N$_2$}&{\small 29957}&{\small 4280}&{\small 881}&{\small 46052}&{\small 6579}&{\small 1354}\\
\hline
\end{tabular}
\label{penetration}
\end{table}

A larger column density of solid D$_2$O ice is needed to reach 95 and 99\% of the total photon absorption with respect to the solid H$_{2}$O values, due to its lower VUV absorption 
cross section. The same holds for solid CD$_3$OD and CH$_3$OH, with the exception of the Avg. value, which is higher in the solid CD$_3$OD sample. For solid $^{13}$CO$_2$ a larger 
column density is needed to reach 95 and 99\% of the total photon absorption with respect to solid CO$_{2}$. In the other hand, within the significant errors associated to their 
VUV absorption cross section measurements, $^{15}$N$_2$ ice seems to be similar or slightly more absorbing than N$_2$. For the first time, we report the VUV absorption cross section 
as a function of photon energy for D$_2$O, CD$_3$OD, $^{13}$CO$_2$, and $^{15}$N$_2$ in the solid phase at 8 K. 

All four molecules present a shift to shorter wavelengths in their VUV spectrum with respect to their corresponding light isotopologues. Deuterated species experience the largest blue-shift among 
the molecules studied. This could be expected from previous works on deuterated species in the gas phase, but the shifts measured in the solid phase were larger in comparison. 
The average and the integrated VUV absorption cross section values are close for the different isotopologues. The relatively small variations between isotopologues may only play a 
minor role in the absorption of VUV radiation in space. 

Large differences were found between the VUV absorption cross section spectra of solid and gas phase species (Papers I, II, and ref. therein; this work). This has important 
implications for the absorption of VUV photons in dense clouds and circumstellar regions. 

There is a clear correspondence between the photodesorption rates measured at different photon energies and the VUV absorption spectrum for the same photon energies. 
This indicates that photodesorption of some ice species like N$_2$ and CO is mainly driven by a desorption induced by electronic transition (DIET) process 
(Fayolle et al. 2011, 2013). Unfortunately, the N$_2$ and $^{15}$N$_2$ ice absorption spectra at photon energies higher than 12.4 eV, where 
photodesorption is efficient, have not been measured. But the low photodesorption rates measured at energies below 12 eV by \cite{Fayolle2} (no more than 4 $\times$ 
10$^{-3}$ molecules per incident photon for $^{15}$N$_2$ ice) is compatible with its low ice absorption cross section, reported here for the same spectral range. In addition, the 
observed photodesorption occurs in the same spectral range where the absorption bands of Fig.~\ref{15N2} are present. The lower photodesorption reported by Fayolle et al. 2013 
for the Ly-$\alpha$ 
wavelength at 121.6 nm, 1.5 $\times$ 10$^{-3}$ molecules per incident photon, coincides with a low absorption in Fig.~\ref{15N2}, and the maximum in the photodesorption occurs 
approximately at $\sim$ 135 nm, where the most intense absorption band is present, see Fig.~\ref{15N2}. The photodesorption rate per absorbed photon in that range, 
$R^{\rm abs}_{\rm ph-des}$, can be estimated as follows
\begin{equation}
 R^{\rm abs}_{\rm ph-des} = \frac{I_0}{I_{abs}} \; R^{\rm inc}_{\rm ph-des}
\end{equation}
where
\begin{eqnarray}
I_{abs} &=& \displaystyle\sum\limits_{\lambda_i}^{\lambda_f} \quad I_0(\lambda) - I(\lambda) = \displaystyle\sum\limits_{\lambda_i}^{\lambda_f} \quad I_0(\lambda)(1 - e^{- \sigma(\lambda) N}) \nonumber
\end{eqnarray}
and $I_0$ is the total photon flux emitted (Fayolle et al. 2013 reports 3-11.5 $\times$ 10$^{12}$ photons cm$^{-2}$ s$^{-1}$, in our experiments this flux is about 2.0 $\times$ 
10$^{14}$ photons cm$^{-2}$ s$^{-1}$), $I_{abs}$ is the total photon flux absorbed by the ice, $I_0(\lambda)$ is the photon flux emitted at wavelength $\lambda$, $\sigma(\lambda)$ 
is the VUV absorption cross section at the same wavelength, and $N$ is the column density of the ice sample. $R^{\rm inc}_{\rm ph-des}$ corresponds to a photodesorption rate of $\leq$ 4 $\times$ 
10$^{-3}$ molecules per incident photon in the spectral range below 12.4 eV for $N$ = 60 $\times$ 10$^{15}$ cm$^{-2}$ (60 monolayers) from \cite{Fayolle2}, while 
the average absorption cross section for $^{15}$N$_2$ ice that we measured in that range is $\sigma$ = 8.7 $\pm$ 1.9 $\times$ 10$^{-21}$ cm$^{2}$. The resulting photodesorption rate 
is thus quite high, $R^{\rm abs}_{\rm ph-des}$ $\leq$ 7.7 molecules per absorbed photon, meaning that a very small fraction of the incident photons are absorbed in the ice but each 
absorbed photon led to the photodesorption of about 7.7 molecules on average (this in fact is the maximum value because \cite{Fayolle2} measured photodesorption 
rates, $R^{\rm inc}_{\rm ph-des}$, that \emph{do not exceed} 4 $\times$ 10$^{-3}$ molecules per incident photon). 

In the case of CO ice deposited also at 15 K, it was found that only the photons absorbed in the top 5 monolayers led to photodesorption with a rate of 2.5 CO molecules per absorbed 
photon in those 5 monolayers (based on Mu\~noz Caro et al. 2010, but using an average cross section of CO ice of 4.7 $\pm$ 0.4 $\times$ 10$^{-18}$ cm$^2$ adapted from 
Paper I). This value for CO ice is about 3.1 times lower than the maximum estimated above for the 60 ML of the $^{15}$N$_2$ ice experiment of \cite{Fayolle2}. A more 
direct comparison between N$_2$ and CO ice photodesorption could be made if the number of N$_2$ monolayers closer to the ice surface that truly contribute to the photodesorption was 
known (in the case of CO ice these are $\sim$ 5 monolayers, this value has not been estimated for N$_2$ ice and therefore the values of $R^{\rm inc}_{\rm ph-des}$ and 
$R^{\rm abs}_{\rm ph-des}$ correspond to the total ice column density of 60 monolayers in the experiment of Fayolle et al. 2013). With this uncertainty still remaining, 
we can conclude that if the VUV absorption cross section of each specific ice composition is taken into account, it is possible to know what is the efficiency of the photodesorption 
per absorbed photon; in the case of N$_2$ and CO, for VUV photon energies that do not lead to direct dissociation of the molecules in the ice, these values are higher than unity. 
The values of $R^{\rm abs}_{\rm ph-des}$ \textgreater 1 and the fact that the photons absorbed in ice monolayers deeper than the top monolayers (up to 5 for CO) can lead to a photodesorption 
event, indicate that the excess photon energy is transmitted to neighboring molecules in the ice within a certain range (this range may correspond to about 5 monolayers in the case 
of CO ice, e.g., Rakhovskaia et al. 1995; \"Oberg et al. 2007, 2009; Mu\~noz Caro et al. 2010); if a molecule on the ice surface receives sufficient energy, it may photodesorbs 
(Mu\~noz Caro et al. 2010).

It should also be noted that ice photodesorption experiments performed with a continuum emission source (like the MDHL), mimicking the secondary VUV field in dense cloud interiors, 
can lead to photodesorption rates that are intrinsically different from those obtained in experiments using a monochromatic source (generally provided by a synchrotron beam), we 
refer to \cite{Chen2} for the case of CO ice photodesorption. 

This work, along with Papers I and II, provides essential data to attempt a more quantitative study of VUV absorption of molecules forming ice mantles, and the photon processes 
involved: photo-processing leading to destruction of molecules and formation of new species, and photo-desorption of molecules in the ice that are ejected to the gas phase. 

\section*{Acknowledgments}

This research was financed by the Spanish MICINN under projects AYA2011-29375 and CONSOLIDER grant CSD2009-00038. This work was partially supported by NSC 
grants NSC99-2112-M-008-011-MY3 and NSC99-2923-M-008-011-MY3, and the NSF Planetary Astronomy Program under Grant AST-1108898.

\bsp

\label{lastpage}

\end{document}